\documentclass{aastex}

\shorttitle{Shell Structure in the Fornax Dwarf Spheroidal Galaxy}
\shortauthors{Coleman et al.}

\begin{document}

\title{Shell Structure in the Fornax Dwarf Spheroidal Galaxy}

\author{Matthew Coleman}
\affil{Research School of Astronomy \& Astrophysics, Institute of 
Advanced Studies, The Australian National University, Cotter Road, Weston 
Creek, ACT 2611, Australia}
\email{coleman@mso.anu.edu.au}

\author{G. S. Da Costa}
\affil{Research School of Astronomy \& Astrophysics, Institute of 
Advanced Studies, The Australian National University, Cotter Road, Weston 
Creek, ACT 2611, Australia}
\email{gdc@mso.anu.edu.au}

\author{Joss Bland-Hawthorn}
\affil{Anglo-Australian Observatory, PO Box 296, Epping, NSW 2121, Australia}
\email{jbh@aaoepp.aao.gov.au}

\author{David Mart\'{\i}nez-Delgado}
\affil{Max-Planck-Institut f\"ur Astronomie, K\"onigstuhl 17, D-69117 
Heidelberg, Germany}
\email{ddelgado@mpia-hd.mpg.de}

\author{Kenneth C. Freeman}
\affil{Research School of Astronomy and Astrophysics, Mount Stromlo 
Observatory, Cotter Road, Weston Creek, ACT 2611, Australia}
\email{kcf@mso.anu.edu.au}

\and

\author{David Malin}
\affil{Anglo-Australian Observatory, PO Box 296, Epping, NSW 2121, Australia}
\email{dfm@aaoepp.aao.gov.au}

\slugcomment{Version of 5 November, 2003}
\begin{abstract}
We present initial results from a wide field photometric survey of the 
Fornax dwarf spheroidal (dSph) galaxy.  The focus here is on a small 
overdensity of stars located approximately $17'$ ($670$ pc in projection) 
from the centre of Fornax.  Based on imaging data in both $V$ and $I$ 
bands down to $V \approx 21$, the dimensions of the feature are 
$\sim$1.7$' \times 3.2'$ ($68 \times 128$ pc) with an 
average surface brightness of $25.8$ mag/arcsec${}^2$ in $V$.  A
follow-up analysis using the
deep $B$ and $R$ band data obtained by \citet{ste98} indicates that the 
overdensity is apparently dominated by a relatively young stellar population with an 
age $\approx 2$ Gyr, though the current data do not rule out the presence of older stars.  Our preferred interpretation is 
that the overdensity represents shell structure in the Fornax dSph galaxy, 
a phenomenon previously unseen in dwarf galaxies.  The shell may be the 
remnant of a merger with Fornax of a smaller, gas-rich system  
that occured approximately 2 Gyr ago.  The comparitively recent star formation
seen in some dSph and dE galaxies may then also be the result of infall of
small relatively gas-rich systems.
\end{abstract}


\section{Introduction}

Dwarf spheroidal (dSph) galaxies provide important clues to our understanding 
of the universe. For example, the high mass-to-light ratios found in 
these systems make them excellent laboratories for the study of dark 
matter.  Further, due to the relatively large number of dwarf galaxies, they 
are useful tracers of matter (both luminous and dark) in the local universe.  
At present, the most widely accepted model describing the formation of 
structure in the 
universe is the Cold Dark Matter (CDM) paradigm, based on hypothetical 
non-baryonic particles \citep{pee93,pea99}.  CDM predicts that relatively 
small-scale clumps of dark matter formed from fluctuations in the density 
spectrum of the early universe.  Large structures were then formed through 
a process of hierarchical merging.  According to this theory, the initial 
dark matter haloes had masses of $\sim$10$^7$ M${}_{\odot}$ \citep{may02,
moo99}, somewhat smaller than the masses of Galactic dSphs ($2 \times 
10^7$ M${}_{\odot}$ to $7 \times 10^7$ M${}_{\odot}$, e.g.\ \citet{mat98}).  
Consequently, the inference from the simulations is that the current
Galactic dSphs have also formed from accretion events in the early
Universe.  However, high resolution simulations of individual dark matter halos (Power 2003, personal communication) indicate that late infall is negligible in typical
dwarf halos.  Therefore, discounting a fleeting encounter 
with a dark clump in the Galactic halo, dwarf galaxies are not expected to
show any significant substructure.
Nevertheless, substructure is observed in at least one Galactic dSph,
namely Ursa Minor (e.g.\ \citet{oa85,dmd01,bel02,pms03,kwg03}), though in
this particular case the origin of the substructure may lie with the tidal 
field of the Galaxy.

Although CDM accurately reproduces the large-scale structure of the
 universe, its predictions do not match observations of small scale regions 
with a high contrast in density.  For example, 
the mass distribution of low surface brightness (LSB) galaxies inferred from 
their rotation curves do not have the steep density cusp at the core expected 
from CDM models (e.g.\ \citet{wel03}).  In addition,
the number of dwarf galaxies surrounding our Galaxy is an order of magnitude 
smaller than the mass spectrum predicted by CDM (e.g.\ \citet{moo99}).  
Further, in the CDM picture, the 
Galaxy's halo was built up through mergers of low-mass systems such as dwarf 
galaxies.  However, the abundance patterns observed in present day dwarfs 
appear to be distinct from those exhibited by Galactic halo field stars  
\citep{she01,tol03}.  Indeed, age, chemical and dynamical inconsistencies 
make currently existing dwarfs 
unlikely to be the major building blocks of large systems such as spiral 
and elliptical galaxies (e.g.\ \citet{tos02}).

The 
Fornax dwarf spheroidal galaxy is the brightest object of this type in the 
Galactic halo ($M_V = -13.2$, from \citet{mat98}).  It lies at a distance of 
$138 \pm 8$ kpc, and has a tidal radius of 
$r_t = 71' \pm 4'$  \citep{ih95}
equivalent to $r_t = 2.85 \pm 0.16$ kpc.  Fornax is one of nine known dwarf 
spheroidal companions of the Milky Way, all of which show little or no 
indication of HI or HII (with the possible exception of Sculptor, see 
\citet{car98,bou03}).  Fornax displays a very complex star formation history, 
beginning at least 12 Gyr ago and extending intermittently until 
$\sim 500$ Myr ago (see \citet{ste98,sav00} and references therein).  
There are also indications, in the form of an excess of stars beyond
the `tidal radius' of the best-fit King model, that the gravitational 
potential of the Galaxy is tidally heating the outer regions of this dSph 
\citep{ih95}.  However, \citet{wal03} find no such excess, albeit for
a larger `tidal radius' ($r_t = 98'$) than found by \citet{ih95}.
Possible asymmetries in the distribution of stars in the inner regions 
of Fornax have been reported by \citet{pwh61}, \citet{esk88} and 
\citet{ste98}.  The latter authors drew particular attention to the 
difference in distribution between the youngest stars and that 
for the bulk of the population.  A measurement of the proper motion of 
Fornax, made using Hubble Space
Telescope data, indicates that this dwarf is currently at or near
perigalacticon \citep{pia02}.

In this paper we examine an overdensity of stars situated 
approximately $17'$ from the centre of the Fornax dSph.  This overdensity 
does not correspond to any previously reported asymmetry in the distribution 
of Fornax stars.  By collecting 
accurate photometry down to the red clump at $V=21$, we have 
been able to determine the surface brightness and dimensions of this 
feature. 
Subsequent analysis of the deep survey data of Fornax provided by 
\citet{ste98} reveals that the overdensity is apparently dominated by a relatively
young main sequence 
stellar population.  We propose that this clump is the first detection 
of shell structure in a dwarf galaxy: such structure, despite being well 
observed in large galaxies, was not previously known in small systems such as 
dwarf galaxies.  The recognition of shell structure in elliptical galaxies 
(e.g.\ \citet{mal80}) provided an important insight into understanding 
the interactions between a large galaxy and a smaller companion.
Phase-wrapped shells in the 
haloes of large galaxies became obvious through the process of photographic 
amplification \citep{mal78} of images of elliptical galaxies (e.g.\ 
\citet{mal83b}).  Subsequent simulations by \citet{her88,her89} indicated
that a large galaxy disrupting a small companion will produce the
observed structure, with some ($\sim10\%$) of the 
material of the companion galaxy distributed in shells around the 
primary.  The level of ordered structure in the shells reveals the type of 
interaction between the two objects: a direct collision will produce highly 
ordered, phase-wrapped shells around the primary galaxy, while a companion 
on a non-radial orbit will produce a confused set of shells.  Observational 
evidence supports this process of shell formation.  In particular, studies 
of the colour of the shells shows that they are 
generally bluer than the colour of the primary galaxy, reflecting
the stellar population of the disrupted lower mass 
companion \citep{pen86,mcg90}.

In the following section we describe our Fornax observations and the data 
reduction techniques.  Section 3 outlines the analysis of 
the data and the results are discussed in section 4.  The conclusions 
are presented in section 5.

\section{Observations and Data Reduction}

Initial deep images of the Fornax dwarf galaxy were derived by combining 
photographically amplified derivatives \citep{mal78, jbh93} from two deep 
plates taken with the UK Schmidt Telescope.  The original exposures (J5462 
and J8297) were made on hypersensitised Kodak IIIaJ emulsion with a Schott 
GG395 filter defining a broad passband from 395 to approximately 540nm.  The 
shell-like structure is readily visible on the derivatives made from both plates 
and is undoubtedly real.  In Fig.\ \ref{fornax_malin}, the two 
photographically-amplified derivatives have been photographically combined 
to improve the signal-to-noise in the final image.  The feature was further
confirmed by inspecting the Digitised Sky Survey, and it is also present
in the APM catalogue \citep{mes96}.  This prompted further analysis using 
CCD data.  

CCD images were obtained with the Siding Spring Observatory 1 metre telescope 
using the Wide-Field Imager (WFI) in October/November 2002.  WFI comprises 
eight 4096 $\times$ 2048 CCDs arranged in a $2 \times 4$ mosaic to give a 
total format of 8192 $\times$ 8192 pixels.  This gives a field-of-view 
of 52 $\times$ 52 arcmin at a scale of $0.38''$ per pixel.  The field
centre was placed at the coordinates $\alpha=02^h 41^m 34.2^s$, 
$\delta=-34^{\circ} 48' 06''$ (J2000.0), which lies approximately $31.8'$ 
southeast from the centre of Fornax.  Thus, the field includes the central 
region of Fornax.  We recorded images in both $V$ and $I$ with exposure 
times of $6 \times 600$s and $6 \times 480$s, respectively, under good 
conditions (FWHM $\approx 1.5''$).  These data form a small part of our 
large-scale photometric study of the Fornax dSph galaxy.

Image reduction was accomplished using standard IRAF routines.  The bias 
images and overscan region allowed the pedestal current to be subtracted 
from all object and flat field exposures.  Each object image was divided 
by the appropriate combined flat, accounting for variations in CCD 
pixel sensitivity.  Fringing effects in the $I$-band are negligible.  After registering and combining the images for each 
filter, instrumental magnitudes were measured using the DAOPHOT 
\citep{ste87} program within IRAF, which allows the point-spread function 
to be determined interactively.  The full-width half-maximum values for 
the PSF were 4.8 pixels ($1.8''$) and 3.5 pixels ($1.3''$) in $V$ and $I$, 
respectively.

The magnitudes and corresponding colours were calibrated 
using \citet{gra82} and \citet{lan92} standard fields observed across the CCDs of the WFI mosaic throughout 
the run.  The extinction coefficients adopted were $k_V=0.16$ and $k_I=0.086$ from \citet{sung00}.  As an example, the
transformation equations for CCD 7 on the nights the data discussed
here were obtained are:
$$
V-v=24.26+0.006(V-I),
$$
$$
I-i=24.79-0.014(V-I)
$$

\noindent where $v$ and $i$ represent extinction corrected instrumental magnitudes using a photometry offset of 23.5.  The rms scatter about these relations are 0.025 and 0.027 mag, respectively.  The equations for CCD 8 are similar.

We present the results here for CCDs 7 and 8 which occupy the north-west 
quarter of the WFI mosaic. These CCDs contain the 
centre of Fornax and the overdense region as shown in Fig.\ \ref{F17V_78}. 
The photometric errors used are those returned by the DAOPHOT program.

\section{Results}

The first question to be addressed is whether the feature might be the
result of differential extinction in the the Fornax field.  We approached
this question in two ways.  First, we chose two regions, each 
$\sim$10$\arcmin$ in size.  One was located near the centre of Fornax and 
the other centred in the vicinity of the feature.  Using the extinction 
maps of \citet{sch98}, the difference in $E(B-V)$ between these two regions 
was investigated and  found to be less than 0.01 mag.
Minor changes in the positions of the regions do not alter this result.
Second, we inspected the IRAS 100$\mu$m images of this part of the sky and
found that there is no `cirrus' structure in the vicinity of the Fornax 
feature.  We conclude therefore that the feature is not a consequence of
differential extinction.

We now consider the region of the feature in more detail.  It is 
centred approximately $17'$ ($670$ pc in projection) 
from the centre of Fornax at coordinates 
$\alpha=02^h 40^m 28.5^s$, $\delta=-34^{\circ} 42' 33''$ (J2000.0).
The physical characteristics were determined by convolving
the $V$-band image with a circular Gaussian kernel 
using the IRAF subroutine {\em gauss} and a standard deviation of 
$\sigma=30$ pixels (11.3$\arcsec$).  This $\sigma$ value was chosen to merge small-scale structure within the feature, while allowing us to determine its overall dimensions.  Defining a sampling resolution of 
$x_{res}=y_{res}=64$ pixels (24$\arcsec$), we produced a contour image 
using the subroutine {\em contour} in 
the IRAF package.  The feature is clearly detected in this convolved
image.  It is almost rectangular in 
shape with approximate dimensions $1.7' \times 3.2'$ ($68 \times 128$ 
parsecs at the distance of Fornax). The feature major axis
has a position angle of $41^{\circ} \pm 2.5^{\circ}$ measured eastwards
from north.  This agrees closely with the position angle of the major axis
of Fornax, which \citet{ih95} give as $41^{\circ} \pm 1.0^{\circ}$.  
In other words, the 
major axis of the feature is parallel to the major axis of Fornax. 
The feature also lies very close to the {\em minor} axis of Fornax.
According to \citet{pia02}, the proper motion of Fornax is also approximately
in the direction of the galaxy's minor axis.

Using the IRAF routine {\em polyphot}, we next calculated the average 
surface brightness of a typical region at the same (ellipticity-corrected) 
distance from the centre of Fornax as the feature.  This is  $26.1 \pm 0.1$ 
mag/arcsec${}^2$ in $V$.  The feature, however, has an average 
surface brightness of $25.8 \pm 0.1$ mag/arcsec${}^2$ in $V$, indicating an 
increased stellar density of
approximately 30\%.  The average surface brightness and the 
area of $1.7' \times 3.2'$, together with the background, then correspond to 
a total magnitude for the excess density of 
$m_V = 16.6 \pm 0.6$.  Assuming the stellar population of the overdensity is at 
the same distance as Fornax ($(m-M)_V=20.80$ for $E(B-V)=0.03$ \citep{mat98}), 
the total absolute visual magnitude is $M_V \approx -4.2 \pm 0.6$ and the visual 
luminosity is $L_{V} \approx 4 \pm 2 \times 10^3 L_{\odot}$.  For comparison, the faintest
of the five Fornax globular clusters, Fornax 1, has $M_V = -5.4$ and a limiting radius of $1.25'$ \citep{mg03}.  This cluster is readily visible in the NW of Fig.\ \ref{fornax_malin} and the difference in size compared to the feature is apparent.  Within the Galactic halo, globular clusters of comparable absolute magnitude to the feature include Pal 12 ($-4.5$) and Pal 13 ($-3.7$) \citep{har96}.

In Fig.\ \ref{cmd_region} we show a colour-magnitude diagram (CMD) for Fornax 
using the two CCDs in the northwest quadrant of the WFI mosaic.
As noted above (see also Fig.\ \ref{F17V_78}), these two CCDs cover the 
central parts of Fornax as well as the region of the overdensity.  We were 
able to measure 5195 and 10312 stars, respectively, on these two CCDs.  The
high galactic latitude of Fornax means that contamination from foreground
stars is minimal.  The resulting CMD reaches down to the red clump at 
$V$ $\approx$ 21 with typical magnitude and colour errors less than 0.05 
mag.  We then compared the CMD for the approximately 100 stars in the 
overdense region with that for Fornax as a whole.  This is also shown in 
Fig.\ \ref{cmd_region} where the photometry for the stars in the region
are shown overlaid on the Fornax CMD\@.  No obvious
differences in the CMD distributions are immediately apparent but this is not 
surprising given that approximately two in every three stars in the 
overdense region are members of the Fornax field rather than of the 
overdense region.  A deeper dataset is needed for further analysis.

Fortunately, the deep survey of Fornax conducted by \citet{ste98} is
available, providing a basis for additional analysis of 
the stellar population of the overdensity.  The \citet{ste98} data were 
obtained using the Cerro Tololo 1.5 m telescope, and a 
full description of their data reduction techniques can be found in their
paper.  Whereas our data ends at the red clump of Fornax 
at $V \approx 21$, the Stetson images provide photometry for $119850$ 
stars down to $V \approx 24$ over a $1/3$ degree${}^2$ field, revealing, 
inter alia, a relatively young main sequence.  The overdense region lies at 
the edge of the Stetson field.  

Figure \ref{stetsoncmd} shows the CMD for the entire Stetson dataset 
with the stars from the overdense region superposed.  There are 327 
stars in this region and the excess above the expected star numbers 
at this (ellipticity-corrected) radius is approximately 40\%; a value in
reasonable agreement with that derived from the WFI data.  It is immediately
apparent from Fig.\ \ref{stetsoncmd} that there is a
high concentration of relatively young main sequence stars in the 
overdense region.  We have confirmed this occurrence in two ways.  

First, we 
selected from the Stetson dataset those stars in the main sequence CMD 
region outlined in Fig.\ \ref{stetsoncmd}.  The spatial distribution of
these stars 
is shown in Fig.\ \ref{stetsonxy_cmdsel}.  The overdense region stands 
out prominently in the bottom left of the figure.  Note that we have 
excluded the inner regions of Fornax in making this plot as there are 
potential substantial inter-field completeness variations for stars in
this part of the CMD in the central regions of the dwarf \citep{ste98}.  
The prominence of the overdensity when using relatively blue
stars also explains why the feature is better seen on blue plates.  We
have not conducted a statistical analysis of this map, but it appears
that the overdense region is the largest fluctuation.  A more
detailed analysis, which allows for variable incompleteness, may reveal
additional overdensities.  \citet{ste98}  did note, however, that the younger main sequence (blue) stars, defined as those with $18.5 \le (B+R)/2 \le 22$ (essentially equivalent limits apply for $B$ for these stars which have $B-R \approx 0$) possess ``a clearly flattened
distribution on the sky with the long axis in the E--W direction and an
axial ratio of 3 or 4 to 1''.  This distribution is evident in
Fig.\ \ref{stetsonxy_cmdsel} as the excess density of points to the East and West of the excluded region; see also Fig. 12 of \citet{ste98}.

Second, in order to make an unbiased assessment of the stellar population
belonging to the overdense region, we have endeavoured to generate a
background-subtracted CMD for the region.  Two complimentary methods were constructed for this CMD subtraction process.  For the first, we defined a 
CMD density function $\Phi$, generated by dividing the Fornax CMD into a grid 
of $34 \times 34$ cells in colour and magnitude and then counting the
stars per cell.  The cells were equally spaced, extending through the domain $-0.4 \le (B-R) \le 3.0$ and the range $19.15 \le B \le 23.95$.  Each cell covered an area $0.1 \times 0.15$ mag${}^2$.  An estimate of the CMD for the overdense region was then
determined by subtracting the `average' Fornax CMD density function 
$\Phi_{Fornax}$ from that for the overdense region.  That is, 

\begin{displaymath}
\Phi_{feature} = \Phi_{region} - \Phi_{Fornax}.
\end{displaymath}

\noindent To avoid the introduction of errors in the function $\Phi_{Fornax}$ 
by any possible radial change in the stellar population of Fornax, we 
chose stars only at the same (ellipticity-corrected) radius 
from the centre of Fornax as the overdense region.  In other words, we used
an annulus 
with the same ellipticity-corrected radius and width as the overdense 
region to select the Fornax stars for our average CMD for the dSph.  
The sample for $\Phi_{Fornax}$ was then normalised by reducing the area of 
the annulus to be the same as that of the overdense region.  This gave
the number of stars in an average region at the same radius 
as the overdense region as $\sim$226, while the \citet{ste98} data
for the overdense region contains 327 stars.  This leads to a further 
estimate of the feature as a $\sim$45\% overdensity, again consistent with 
the previous estimates above.

We then calculated the difference between $\Phi_{region}$ and 
$\Phi_{Fornax}$,  and the resulting CMD density function for the overdense 
region is shown in Fig.\ \ref{subtractioncmd}.  The greyscale shown ranges
from zero (white, indicating no difference between the overdense region and 
Fornax) to 500 (black, indicating a difference of 500 stars/mag${}^2$ at that 
location in the CMD) so that the darker regions in 
Fig.\ \ref{subtractioncmd} indicate higher levels of confidence in the
presence of overdensity members at that colour-magnitude value.
Once again the CMD of the feature appears dominated by a relatively young 
main sequence population, consisting of approximately 80 stars.  There is also some indication of the presence
of red clump and red giant branch stars.

The second subtraction method used the `average' Fornax field CMD defined above.  We constructed a series of field CMDs by using a Monte Carlo alogrithm applied the density function, $\Phi_{Fornax}$.  Each field CMD contained 226 stars, the average number of stars in a region with the same area and distance from the Fornax centre as the overdensity.  We then subtracted each of these trial field CMDs by `cancelling out' the nearest star in the CMD of the overdense region.  A typical outcome from this process is shown in Fig.\ \ref{montecarlo}.  In particular, we find that the occurrence of stars in the region $B \sim 22.5 - 23.5$ and $(B-R) \sim 0.3$ in the subtracted CMDs is a robust result of this process.

To estimate the age of the stars in the overdense region we have used
isochrones calculated from the Yonsei-Yale web-based release 
\citep{yi01,kim02}.  Shown in Figs.\ \ref{subtractioncmd} and \ref{montecarlo} are isochrones 
for ages of 1, 2 and 3 Gyr and an assumed metallicity of [Fe/H] = --1.0.  
Given that the data are increasingly incomplete below B $\approx$ 23.5
\citep{ste98}\footnote{\citet{ste98} actually state that they regard the
data as `essentially complete' to well below [$V=(B+R)/2$] magnitude 23.
For main sequence stars with $B-R \approx 0.4$, $V$=23 corresponds 
to $B$=23.2 from which we assumed that the data are increasingly incomplete
below $B \approx 23.5$.  It is also worth noting, as Figure 2 of \citet{ste98} illustrates, that the region of the overdensity received less total exposure time than the central regions of the dwarf galaxy.}, the isochrones in Figs.\ \ref{subtractioncmd} and \ref{montecarlo} suggest the overdense region is dominated by a stellar population with an age
of $\sim$2 Gyr for this metallicity.  However, the limiting magnitude of the \citet{ste98} data in this region makes it difficult to decide whether the overdense region also contains an older population.  The metallicity assumed was chosen
on the basis of the fit to the CMD, and we note that \citet{tol01} give
the mean metallicity of Fornax as [Fe/H] $\approx -1.0$.  \citet{tol03}
argue for an age-metallicity relation in Fornax and assign an age of
$\sim$2 Gyr to the Fornax giant M21 whose iron abundance, measured from
a UVES high dispersion spectrum, is [Fe/H] =  --0.7.  However, the use of
higher metallicity isochrones produces a less satisfactory agreement in
colour at the main sequence, and does not substantially alter the age
estimate.  Deeper imaging
specific to the overdense region would aid substantially in 
constraining the age (or age range) of its stellar population.  

\section{Discussion}

We turn now to the interpretation of the overdense region.
Our current understanding of dSph galaxies indicates they are complex
systems, with star formation histories extending back to the earliest 
times (\citet{mat98} and references therein).  These systems may be the 
darkest in the universe, with mass-to-light ratios ranging from 
$M/L \sim 5$ for the higher luminosity dwarfs such as Fornax and Sculptor, 
to $M/L \sim 100$ for systems such as Draco and Ursa Minor \citep{mat98}.  
According to CDM theory, the presence of such a dominant dark halo should 
inhibit the formation of significant substructure, barring strong encounters 
with objects of a comparable or higher mass (e.g.\ the Sagittarius and 
NGC~205 dwarf galaxies are being tidally stripped by their host galaxies).  
Therefore, we do not expect the feature to be intrinsic structure in the
dSph.  However, \citet{kwg03} have recently argued that the observed
morphological substructure in the Ursa Minor dSph could be a primordial 
artifact.  From their models, the primary requirements for a dynamically 
cold clump to survive are that the dark matter potential possess a constant
density core, and that core size be substantially larger than the orbit: 
substructure is rapidly destroyed in the cuspy dark matter potential 
predicted by CDM or for orbits larger than the core radius \citep{kwg03}.  
Nevertheless, the Fornax overdense region is unlikely to be a primordial 
survivor (i.e.\ formed at early epochs) as it contains stars as young as 
a few Gyr (cf.\ Figs.\ \ref{subtractioncmd} and \ref{montecarlo}).

An initially plausible interpretation of the overdense region is that
it is a consequence of tidal interaction between Fornax and the Milky
Way Galaxy.  For example, recent studies of the Ursa Minor dSph (UMi) 
by \citet{dmd01,bel02,pms03} and \citet{kwg03} have revealed a strongly 
disturbed structure for this dSph.  This is frequently interpreted as the 
result of tidal interactions with the Galaxy.  However, this is unlikely to 
be the origin of the overdense region in Fornax, for two reasons.  First, the 
galactocentric distance and mass of Fornax are considerably larger than
those of Ursa Minor and thus Fornax is subject to substantially lower
tidal forces.  
Second, and more compelling, is the fact that the overdense region has a 
stellar population that appears to be distinct from that for the bulk of Fornax.  If the 
feature had a tidal origin, the expectation would be that the feature 
population should be the same as that for the dwarf as a whole, and this is
not the case.

Another possible interpretation is that the overdense region represents the
remnant of a disrupted stellar cluster, which presumably was a few Gyr
in age.  As noted above, the \citet{kwg03} models illustrate that a clump formed in this way
could survive if two main criteria are fulfilled.  Firstly, the distribution of the dark matter (DM) in 
Fornax must have an approximately uniform density core (that is, it differs from the standard CDM prediction).  If the core is cusped, then any coherent structure is disrupted within a few orbital periods.  We estimate the orbital period of the clump at this radius to be $\sim 10^8$ years using the Fornax central density and core radius values from \citet{mat98}, considerably less than the inferred age.  Secondly, the core radius of the DM distribution should be at least comparable with the 
distance of the clump from the centre of Fornax.  This latter point is 
possible; the clump lies in projection at about 1.5 r$_{core, stars}$ and the
available velocity dispersion results suggest that the dominant dark mass 
probably has a more extended distribution than the light \citep{mow91}.  The
signature of this interpretation would be that the clump possesses
a single age population, and that it is kinematically cold with a velocity
dispersion substantially less than that of the dwarf.  It would a 
considerable observational challenge to determine if the
overdensity has such as small velocity dispersion.  However, with deeper imaging data it should be possible to determine if the overdensity is dominated by a single age population, as would be expected if it is a disrupted cluster, or contains a mix of populations.  Such deeper imaging observations are planned.  We note, however,
that while on-going star formation has certainly occurred in Fornax, there 
is no evidence for any star clusters in this dwarf other than the five
well known globular clusters, which all have ages comparable
to the globular clusters of the Galactic halo \citep{bon98,bon99} and which 
all show no signs of disruption \citep{mg03,rr94}.  Therefore, it seems unlikely that the clump is the remnant of a disrupted $\sim$2 Gyr old cluster.  We further note that there are {\it no} stellar aggregates (bound or unbound) in any other dSph that have ages of a few Gyr.  Such aggregates that do exist are either very old (such as the Fornax globular clusters or the Ursa Minor kinematically cold clump) or very young.  For example, the Phoenix dSph/dIrr contains an apparent association of young (age $\sim 100$ Myr) stars \citep{dmd99}.  This young association is located towards the centre of the Phoenix dwarf, in contrast to the Fornax feature which is situated significantly beyond the core radius of the dwarf.

Our preferred interpretation is that the overdense region represents 
a `shell' analogous to those seen around many E galaxies: the shells 
arise from orbit phase-wrapping of the debris of a low-mass companion tidally disrupted by a more massive galaxy (e.g.\ 
\citet{her88,her89}).
Such features are
relatively common in more luminous systems, with about $\sim$10\% of all 
E galaxies showing features of this nature \citep{her88}, however such structure has not been previously seen in dE or dSph galaxies.  Our shell 
interpretation is driven by the location and appearance of the Fornax
overdense region -- the location on the minor axis, the elongation 
parallel to the major axis, and the relatively `sharp-edged' nature of the
overdense region are all characteristic of shell features.  Shells around more
luminous E galaxies are also generally bluer than the primary galaxy
\citep{pen86,mcg90} and that is clearly also the case here.  Simulations by 
\citet{her89} predict approximately $10\%$ of the companion mass is distributed as shells around the primary, depending on the mass ratio and collision parameters of the two objects.  If we assume that approximately $1 - 10\%$ of the mass contained in the shells is visible in this clump, we obtain a mass of $\sim 10^6 - 10^7$ M${}_{\odot}$ for the companion object.  This is comparable to the smallest dwarf galaxies of the Local Group.

We propose then that a relatively gas-rich low mass
dwarf galaxy merged with Fornax approximately 2 Gyr ago.  Tidal stirring of gas from the companion produced a new population of stars, which, together with the remnant population of the disrupted system, are distributed as 
shell structure throughout the Fornax dSph galaxy.  Modelling of such a
process may well provide insight as to the location and surface density
of other possible shells to guide further observations.  It may also indicate 
how retention of gas from such a merger process may have led to the unusual 
distribution of the youngest Fornax stars as outlined by \citet{ste98}.  

\section{Conclusion}

We have presented photometric data for a stellar clump located 
$\sim$17$\arcmin$ from the centre of the Fornax dwarf spheriodal galaxy.  
The clump is approximately
68 $\times$ 128 pc in size and is $\sim$40\% more dense than the surrounding 
regions.  It is located on the Fornax minor axis and is elongated in the 
direction of the dSph's major axis.  By removing the Fornax field 
contamination through a CMD density function analysis, we have shown that
the clump appears dominated by a stellar population approximately 2 Gyr in
age.  Our interpretation of this clump is that it is a shell feature
analogous to those frequently seen around more luminous E galaxies.  As such,
it would represent the first detection of shell structure in a dwarf
galaxy.  In this interpretation we propose that a smaller gas-rich system 
merged with Fornax in the relatively recent past ($\sim$2 Gyr ago) and that 
the shell is a consequence of this merger.  Models of such a merger, 
kinematic studies of the shell stars, and further deep imaging of the region 
surrounding this dSph to constrain any additional shell structure, are all 
required to investigate this hypothesis more thoroughly.  It may be that
in many cases the comparitively recent star formation seen in some dSph and dE 
galaxies is also the result of the infall of lower mass, 
relatively gas-rich systems.

\acknowledgments

The authors are grateful to Peter Stetson for making his Fornax data available
for this study.  Matthew Coleman acknowledges the financial support provided
by an Australian Postgraduate Award.  The authors are also grateful to the referee, Dr. Antonio Aparicio, for his helpful comments on the original manuscript.

\clearpage

\plotone{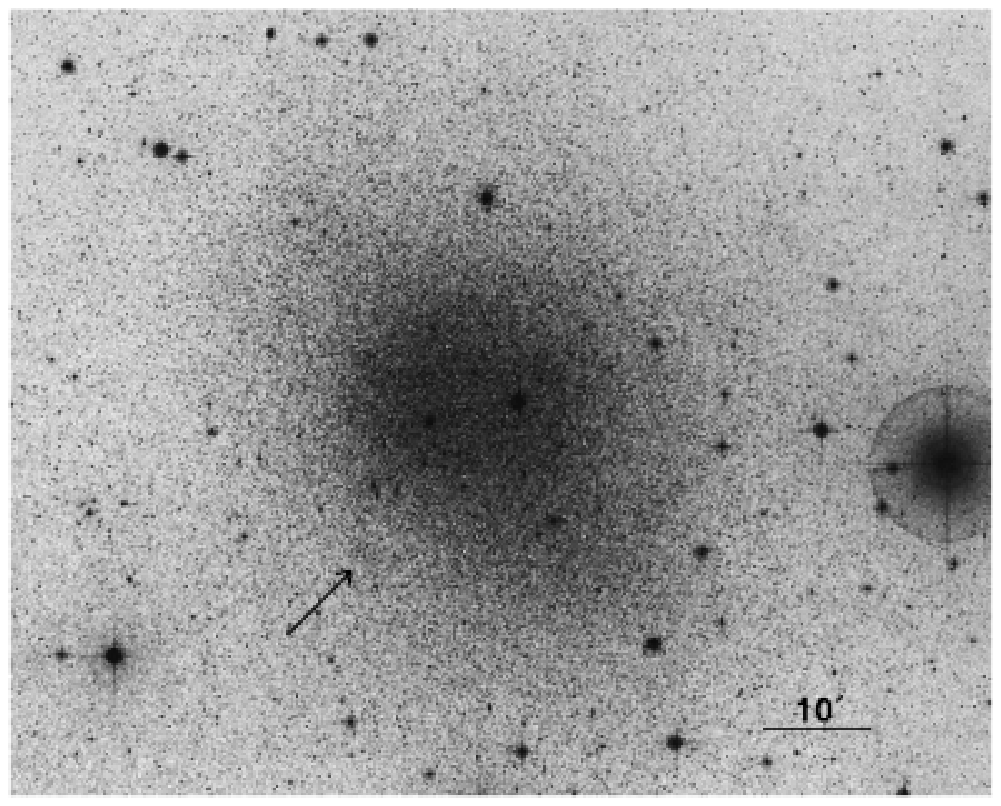}
\figcaption[fornax_malin.eps]{A deep image of Fornax from a combination of 
photographically-amplified derivatives from two UK Schmidt telescope 
IIIaJ plates.  The feature of interest is towards the lower-left of the 
image and is indicated by the arrow.  In this image North is up and East is 
to the left. \label{fornax_malin}}

\plotone{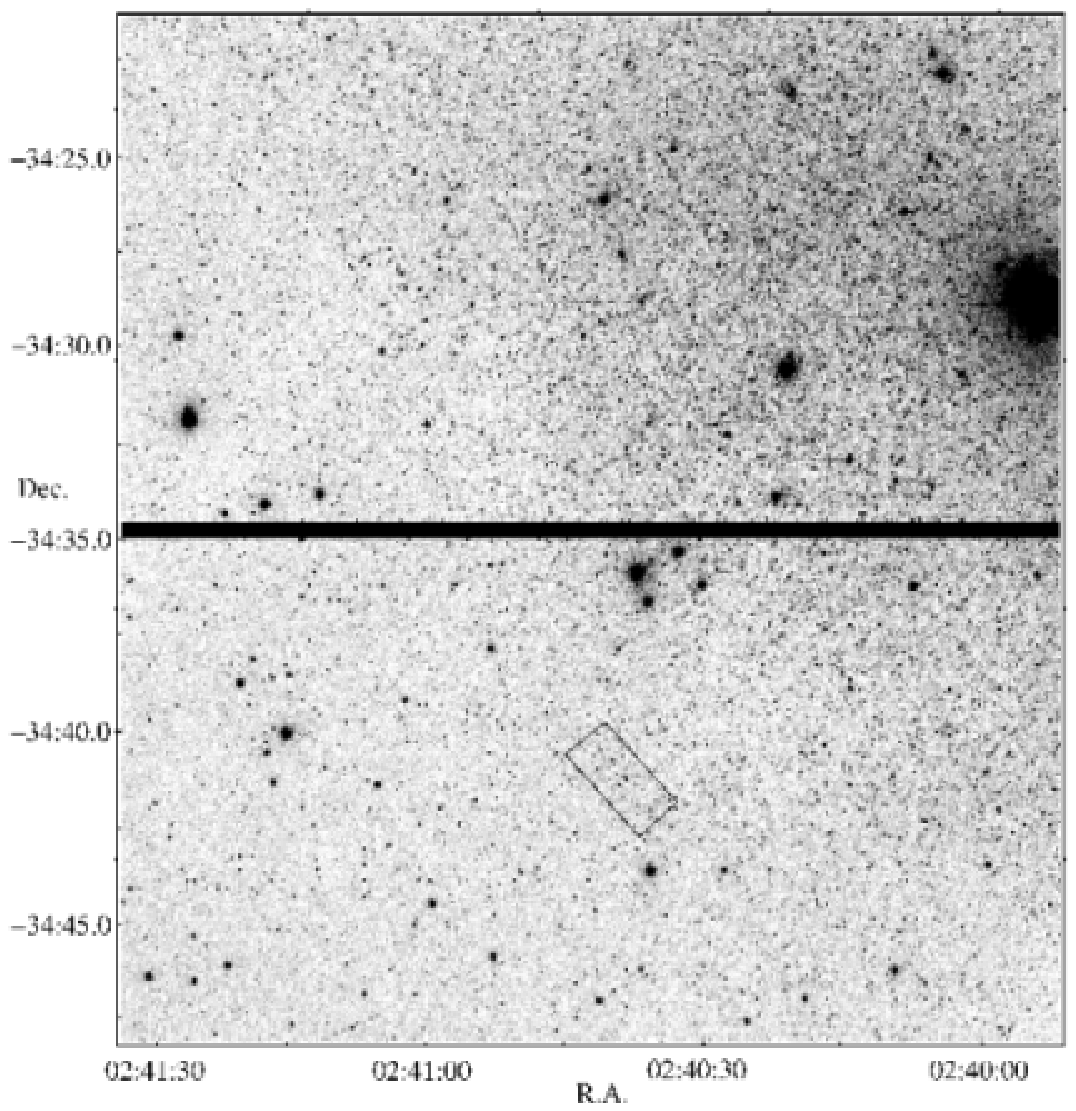}
\figcaption[F17V_78new.eps]{A $V$-band image of Fornax from the Wide-Field 
Imager (WFI) on the 
SSO 1m telescope.  Shown are CCDs 7 (bottom) and 8 (top), which cover an area 
$26' \times 26'$ on the sky.  The overdense feature is the region outlined 
by the rectangle near the centre of CCD 7.  As for Fig.\ \ref{fornax_malin}, 
North is up and East is to the left.  \label{F17V_78}}

\plotone{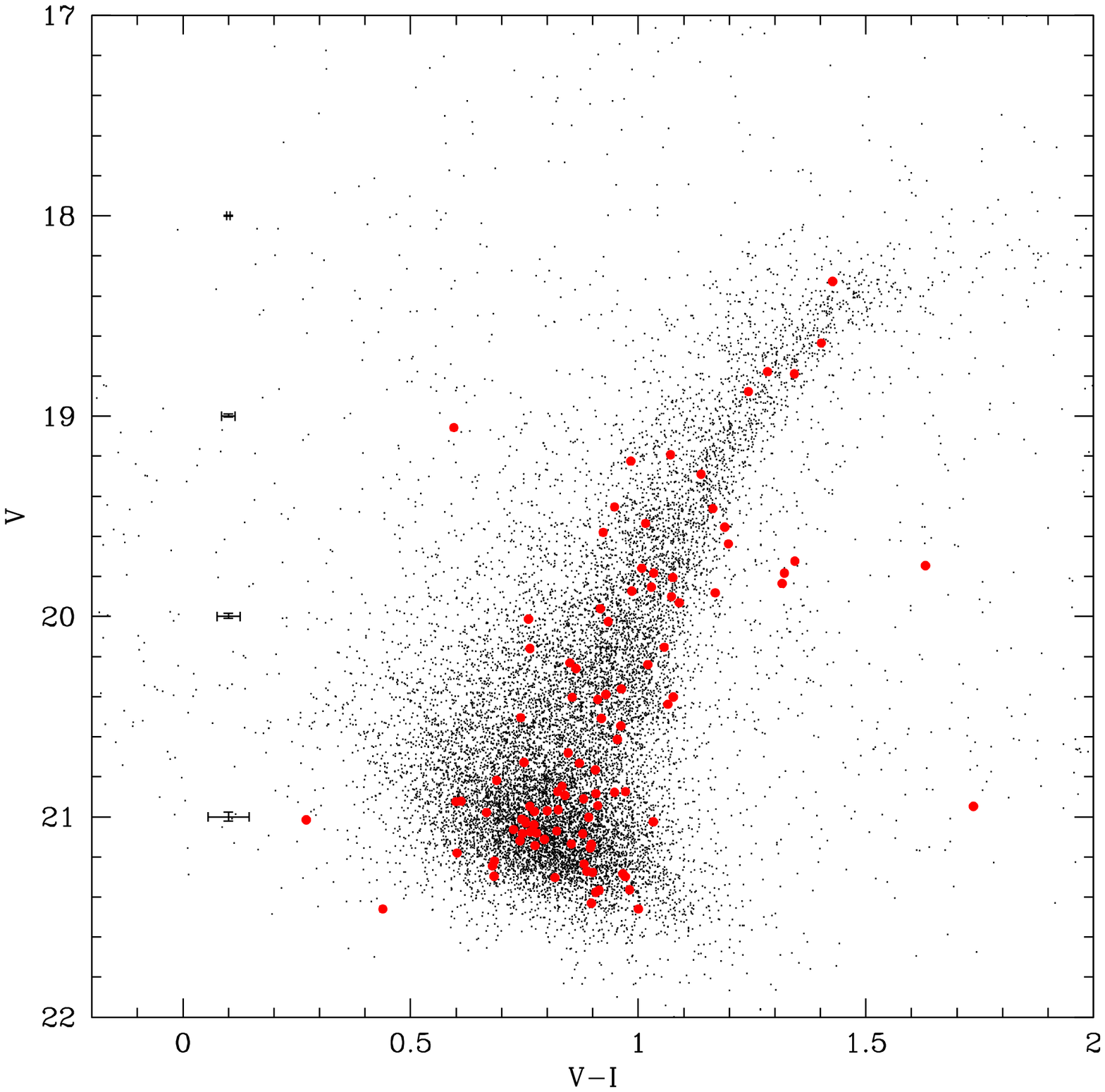}
\figcaption[F17_7cutcmd.eps]{A colour-magnitude diagram for Fornax from WFI 
CCDs 7 and 8 that reaches the red clump at $V \approx 21$.  Overlaid is the 
photometry for the stars in the region outlined in Fig.\ \ref{F17V_78}.  
\label{cmd_region}}

\plotone{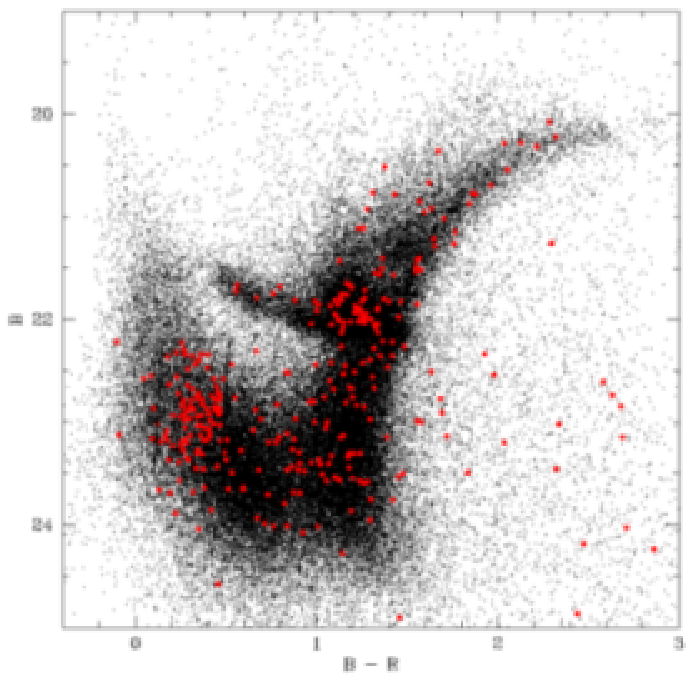}
\figcaption[stetsoncmd_reg.eps]{A colour-magnitude diagram for Fornax from 
the dataset of \citet{ste98}.  Overlaid are the points for the stars that lie
within the boundaries of the overdense region.  A selection box for
main sequence stars is outlined. \label{stetsoncmd}}

\plotone{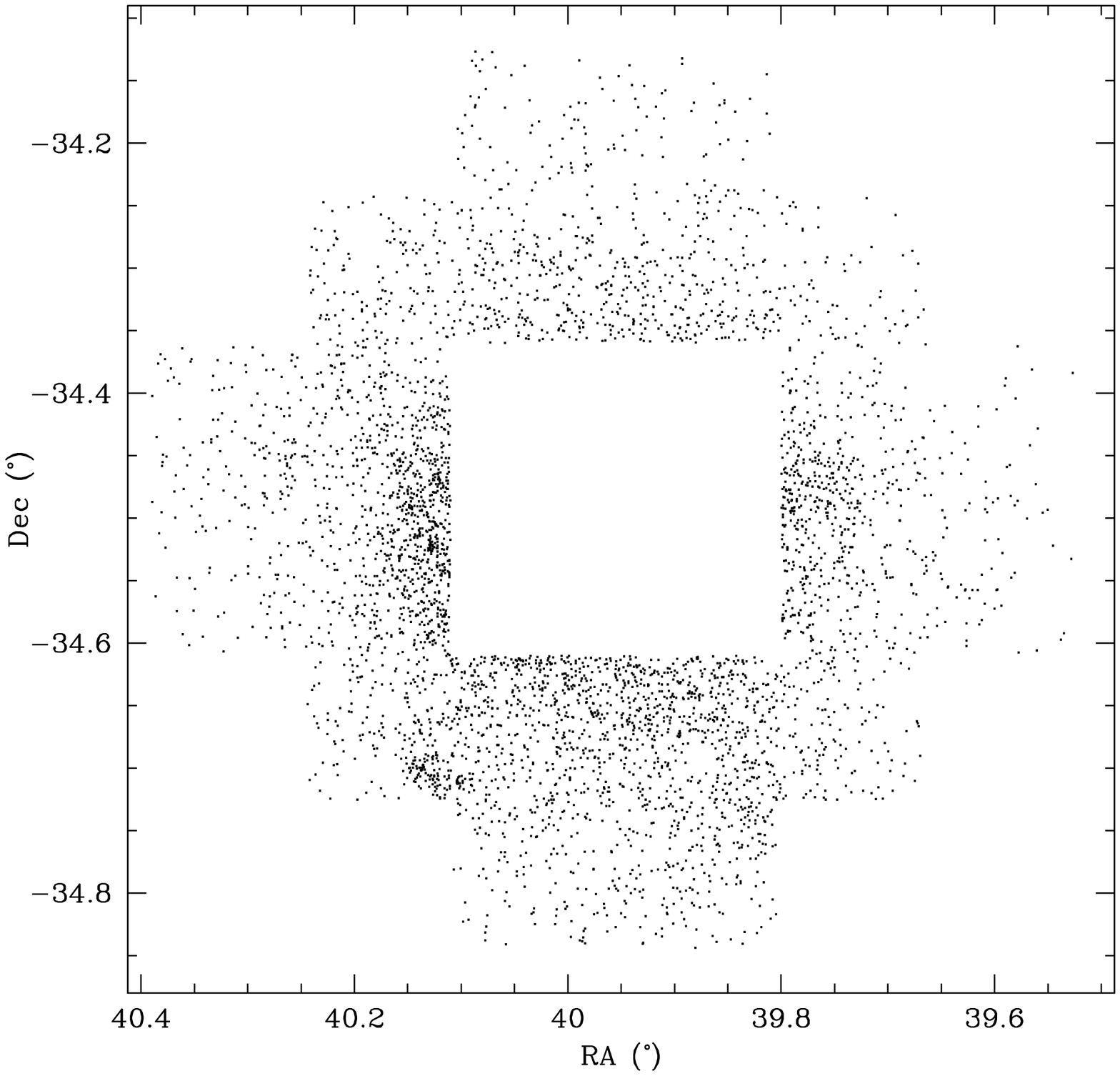}
\figcaption[stetsonxy_cmdsel.eps]{The spatial distribution from the 
\citet{ste98} dataset of main sequence stars in the CMD selection box shown 
in Fig.\ \ref{stetsoncmd}.  The overdense region is clearly visible in the 
lower left of the figure, which has the same orientation as Figs.\ 
\ref{fornax_malin} and \ref{F17V_78}.  The central regions of Fornax 
experience substantial inter-field completeness variations at the 
selected colours and magnitudes, and have been excluded.
\label{stetsonxy_cmdsel}}

\plotone{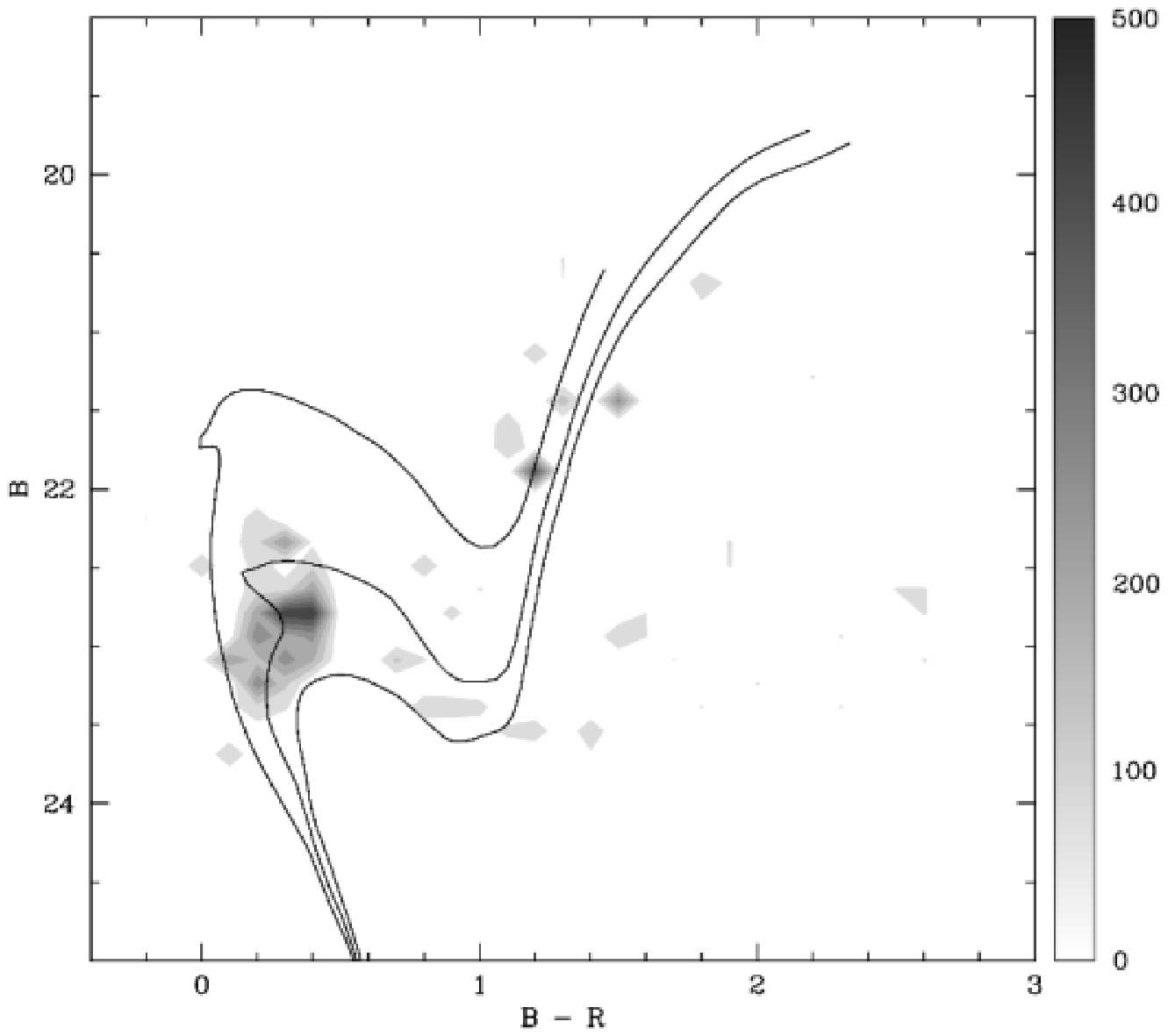}
\figcaption[subtraction_bar.eps]{The background-field-subtracted CMD density 
function for the overdense region, $\Phi_{feature}$.  Overplotted are 
Yonsei-Yale isochrones for ages of 1, 2 and 3 Gyr, and an assumed metallicity
[Fe/H] = --1.0.  The density column at the right hand side defines the 
stellar density in the CMD in stars/mag${}^2$.  The largest arrangement centred at $B \sim 23, (B-R) \sim 0.3$ corresponds to approximately 80 stars.\label{subtractioncmd}}

\plotone{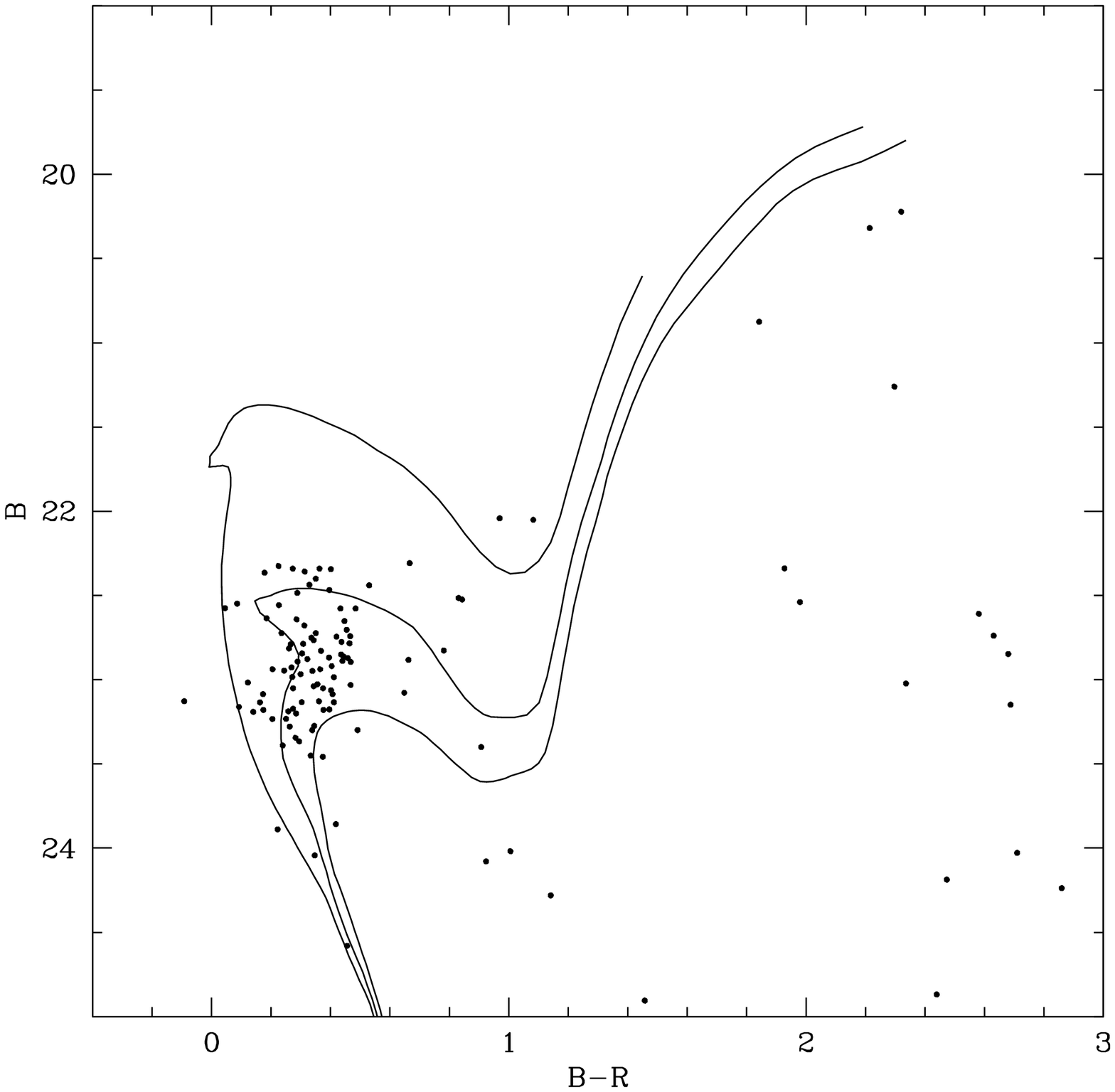}
\figcaption[montecarlo.eps]{Subtracted CMD for the overdense region constructed using a Monte Carlo subtraction technique.  Overplotted are 
Yonsei-Yale isochrones for ages of 1, 2 and 3 Gyr, and an assumed metallicity
[Fe/H] = --1.0.\label{montecarlo}}

\end{document}